\pdfoutput=1
\documentclass[onecolumn]{aastex61}
\usepackage{gensymb}
\usepackage[colorinlistoftodos]{todonotes}
\usepackage[normalem]{ulem}

\shorttitle{A Taxonomic Study of Asteroid Families from KMTNet-SAAO Multi-band Photometry}
\shortauthors{Erasmus et al.}

\begin{document}

\title{\textbf{A Taxonomic Study of Asteroid Families from KMTNet-SAAO Multi-band Photometry}}

\correspondingauthor{Nicolas Erasmus}
\email{nerasmus@saao.ac.za}

\author{N. Erasmus}
\affil{South African Astronomical Observatory, Cape Town, 7925, South Africa.}

\author{A. McNeill}
\affil{Department of Physics and Astronomy, Northern Arizona University, Flagstaff, AZ 86001, USA.}

\author{M. Mommert}
\affil{Lowell Observatory, 1400 W.\ Mars Hill Rd., Flagstaff, AZ, 86001, USA}

\author{D. E. Trilling}
\affil{Department of Physics and Astronomy, Northern Arizona University, Flagstaff, AZ 86001, USA.}
\affil{South African Astronomical Observatory, Cape Town, 7925, South Africa.}

\author{A. A. Sickafoose}
\affil{South African Astronomical Observatory, Cape Town, 7925, South Africa.}
\affil{Department of Earth, Atmospheric, and Planetary Sciences, Massachusetts Institute of Technology, Cambridge, MA 02139-4307, USA.}

\author{K. Paterson}
\affil{Department of Astronomy, University of Cape Town, Cape Town, 7701, South Africa}

\begin{abstract}
We present here multi-band photometry for over 2000 Main-belt asteroids. For each target we report the probabilistic taxonomy using the measured \textit{V-R} and \textit{V-I} colors in combination with a machine-learning generated decision surface in color-color space. Through this method we classify $>$85$\%$ of our targets as one the four main Bus-DeMeo complexes: S-, C-, X-, or D-type. Roughly one third of our targets have a known associated dynamic family with 69 families represented in our data. Within uncertainty our results show no discernible difference in taxonomic distribution between family members and non-family members. Nine of the 69 families represented in our observed sample had 20 or more members present and therefore we investigate the taxonomy of these families in more detail and find excellent agreement with literature. Out of these 9 well-sampled families, our data show that the Themis, Koronis, Hygiea, Massalia, and Eunomia families display a high degree of taxonomic homogeneity and that the Vesta, Flora, Nysa-Polana, and Eos families show a significant level of mixture in taxonomies. Using the taxonomic purity and the degree of dispersion in observed colors for each of the 9 well-sampled collisional families we also speculate which of these families potentially originated from a differentiated parent body and/or is a family with a possible undetermined nested family. Additionally, we obtained sufficient photometric data for 433 of our targets to extract reliable rotation periods and observe no obvious correlation between rotation properties and family membership. 

\end{abstract}


\section{Introduction}
\label{sec:intro}
Main-belt asteroids (MBAs) are the  most abundant designated objects in the solar system, representing approximately $95\%$ of all of the $\sim$800 000 bodies listed in the Minor Planet Center (MPC) catalog\footnote{\url{http://minorplanetcenter.net/}}. This large population of known MBAs therefore lends itself to in-depth statistical investigations to find correlations between orbital and physical properties. One such investigation has led to the identification of asteroid families where the  proper orbital elements of family members display clustering in proper orbital space, suggesting a common collisionally-disrupted parent-body source for family members. Additionally, composition investigations targeting family members have revealed  that in general there is a homogeneity within families \citep{Bus1999}, further supporting the idea behind a common parent or source. Correlation in spectra, color, or albedo within families has also contributed to the fine-tuning of family membership when ambiguity in membership arises from pure orbital considerations \citep{Parker2008}. However, as more and more observing data have become available, certain nuances have come to the foreground which have failed to resolve past questions around asteroid families and in some instances raised new questions. 

For instance, the majority of well-defined families do show consistent spectroscopic properties within a family but there are families that show bi-modal taxonomic distributions. Furthermore, families with homogeneous spectra do also display variations larger than observational uncertainties \citep{Bus1999}. Some explanations that have been proposed include: nested families, i.e., two overlapping families in orbital space that are actually two separate families \citep{Cellino2001}; two colliding parents that had completely different mineralogy; space weathering that modifies spectral shape \citep{Brunetto2005}; a large differentiated parent body having variations in mineralogy \citep{Binzel1993}; or simply incorrectly assigning membership (sometimes referred to as interlopers).

Another contentious topic is the origin of family versus non-family members: did non-family members originate from a quiescent disk with a few large bodies the source of families , or did only a few very large bodies generate both family and non-family members \citep{Dermott2018}.

In this work we present the visible colors for a consistent sample of over 2000 MBAs. The colors are derived from multi-band photometry light curves ranging from $0.35-5.5$ hr (mean: 3.0 hr) which has also allowed us to obtain rotational properties of roughly 1/5th of our targets. By conducting an untargeted survey we have obtained photometry for a sample of MBAs that will be unbiased against location within the main-belt, family or non-family membership, rotation period and amplitude. By analyzing our data set we attempt to decode some of the remaining open questions surrounding asteroid families. We use the entire data set to investigate any differences in the taxonomic distribution of objects belonging to collisional families versus those that do not. We then do a more in-depth analysis of 9 families: Vesta, Flora, Nysa-Polana, Themis, Koronis, Eos, Hygiea, Massalia and Eunomia. We are able to perform a more in-depth study on these families because they had 20 or more confirmed members which were present in our data set.

\section{Observations and Data Reduction}
\label{sec:obs}
\begin{figure}[ht]
	\begin{center}
		\includegraphics[width=0.5\textwidth]{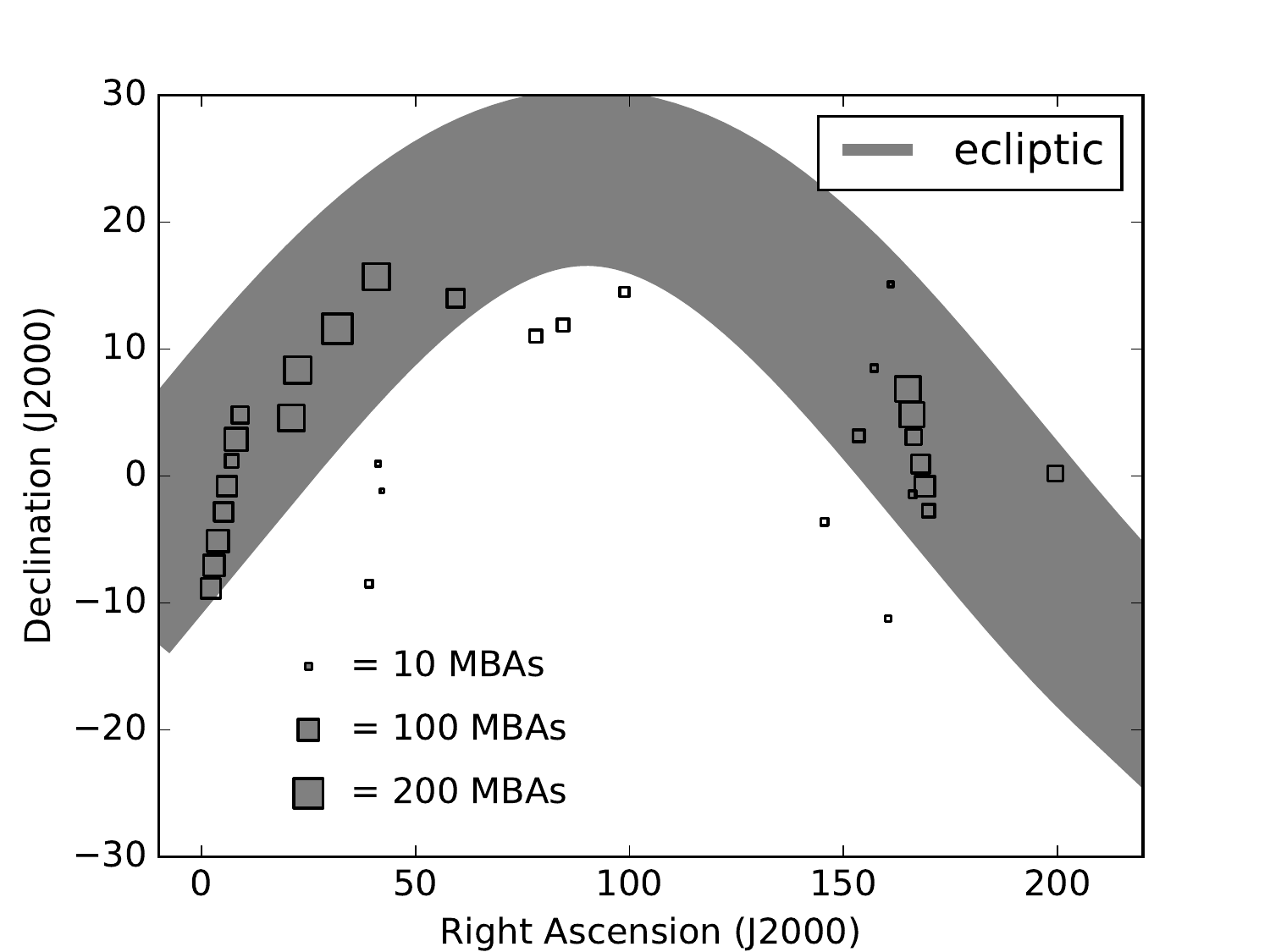}
		\caption{Sky-coordinates of all observed KMTNet fields for the two combined campaigns (hollow squares). The size of the square is proportional to the number of serendipitous MBAs with acceptable SNR extracted in each field. As expected, fields observed on or close to the ecliptic yielded the highest number of MBAs, though some fields on the ecliptic only yielded a few MBAs because unfavorable observing conditions for that specific field meant a large number of MBAs had to be discarded.}
		\label{pointing_coordinates}
	\end{center}
\end{figure}

Observations were performed during two campaigns. The first campaign took place between October 2016 and February 2017 over four observing weeks and was originally intended for a near-Earth asteroid study \citep{Erasmus2017}.  However, the large field of view meant that the fields observed in that study also included $\sim$1000 serendipitously observed MBAs \citep{Erasmus2018}. The second campaign included one observing week in October 2017 and another observing week in February 2018. For the second campaign no specific asteroids were targeted but the observing strategy was to observe close to the ecliptic to maximize the number of MBAs captured in a single field. Approximately 1300 MBAs were observed during the second campaign. Figure \ref{pointing_coordinates} indicates the pointing coordinates for the two combined campaigns.

The telescope used for the two campaigns was the Sutherland, South Africa node of the Korea Microlensing Telescope Network (KMTNet) \citep{Kim2016}. The telescope has a primary mirror 1.6~m in diameter and is fitted with four 9k $\times$ 9k CCDs, mosaicking the 2$\degree$~$\times$~2$\degree$ field of view. Each CCD covers 1$\degree$~$\times$~1$\degree$ of sky with a plate-scale of 0.40 arcsec/pixel. Observations were performed alternating among \textit{V} ($\lambda=550~$nm, $\Delta\lambda=90~$nm), \textit{R} ($\lambda=660~$nm, $\Delta\lambda=140~$nm), and \textit{I} ($\lambda=805~$nm, $\Delta\lambda=150~$nm) filters in the sequence\textit{VRVI}, repeating the sequence continuously for the entire observing duration. The exposure time for each filter was 60~s for campaign one and 120~s for campaign two. Image reductions were performed using the MeerLICHT reduction pipeline (Paterson et al.  in prep) using the KMTNet setting file.

Photometry was extracted from the KMTNet observing data using PHOTOMETRYPIPELINE (PP), an open source Python software package for automated photometric analysis of imaging data, developed by \cite{Mommert2017}. PP also has a feature that enables it to find serendipitously observed asteroids in the image fields using IMCCE's SkyBoT service \citep{Berthier2006}. This feature queries the SkyBoT service for each observed field with a
limiting magnitude that is equal to the 90-th percentile of the
calibrated magnitudes of all sources in the field and a positional
uncertainty of less than 5 pixels. These criteria ensure a reliable
identification of serendipitously observed asteroids in the field.
(See also \cite{Erasmus2017,Erasmus2018} for further details on data reduction and photometry extraction.)

\begin{figure}[ht]
	\begin{center}
		\includegraphics[width=0.5\textwidth]{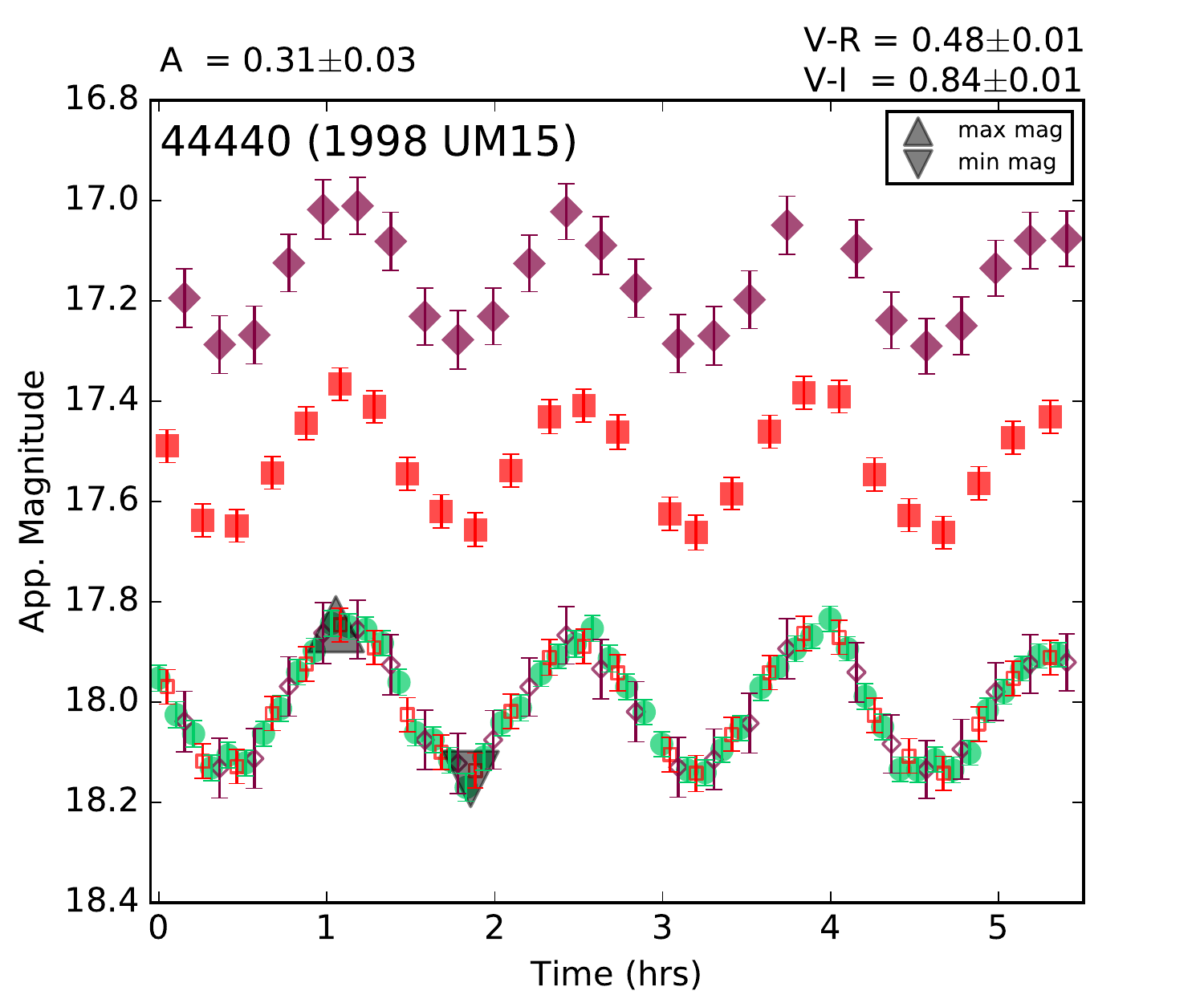}
		\caption{Example photometric data for one of the observed MBAs, 44440 (1998 UM15). The \textit{V}- (green circles), \textit{R}- (red squares), and \textit{I}- (burgundy diamonds) filter data are shown. The hollow symbols in the bottom light curve is the result of adding the determined colors (see Section \ref{sec:color}) to the \textit{R} and \textit{I} data points. Plots and the associated combined (V- together with adjusted R- and I-band) photometric data for all 2276 targets are electronically available for download. (For referee, attached as a zipped file.)}
		\label{photometry_data_example}
	\end{center}
\end{figure} 

\section{Color Calculation and Taxonomy Determination}
\label{sec:color}
To determine the colors of each observed MBA but still accounting for variations in magnitude due to asteroid rotation during an observation, a linear interpolation was performed between adjacent \textit{V}-filter data points and used to obtain a corrected \textit{V} magnitude at times of inter-spaced non-\textit{V} observations. Respective colors were derived by subtracting the non-\textit{V} magnitudes from the interpolated \textit{V} magnitudes. The final determined color is the weighted average (by error in magnitude) of all such possible subtractions in a given observation window. As an example, in the top-right corner of Figure \ref{photometry_data_example} we show the weighted average colors calculated from the \textit{V}, \textit{R}, and \textit{I} photometry data for one of the observed targets. The hollow symbols in the bottom light curve are the result of adding the determined colors to the \textit{R} and \textit{I} data points.

\begin{figure}[ht]
	\begin{center}
		\includegraphics[width=0.5\textwidth]
		{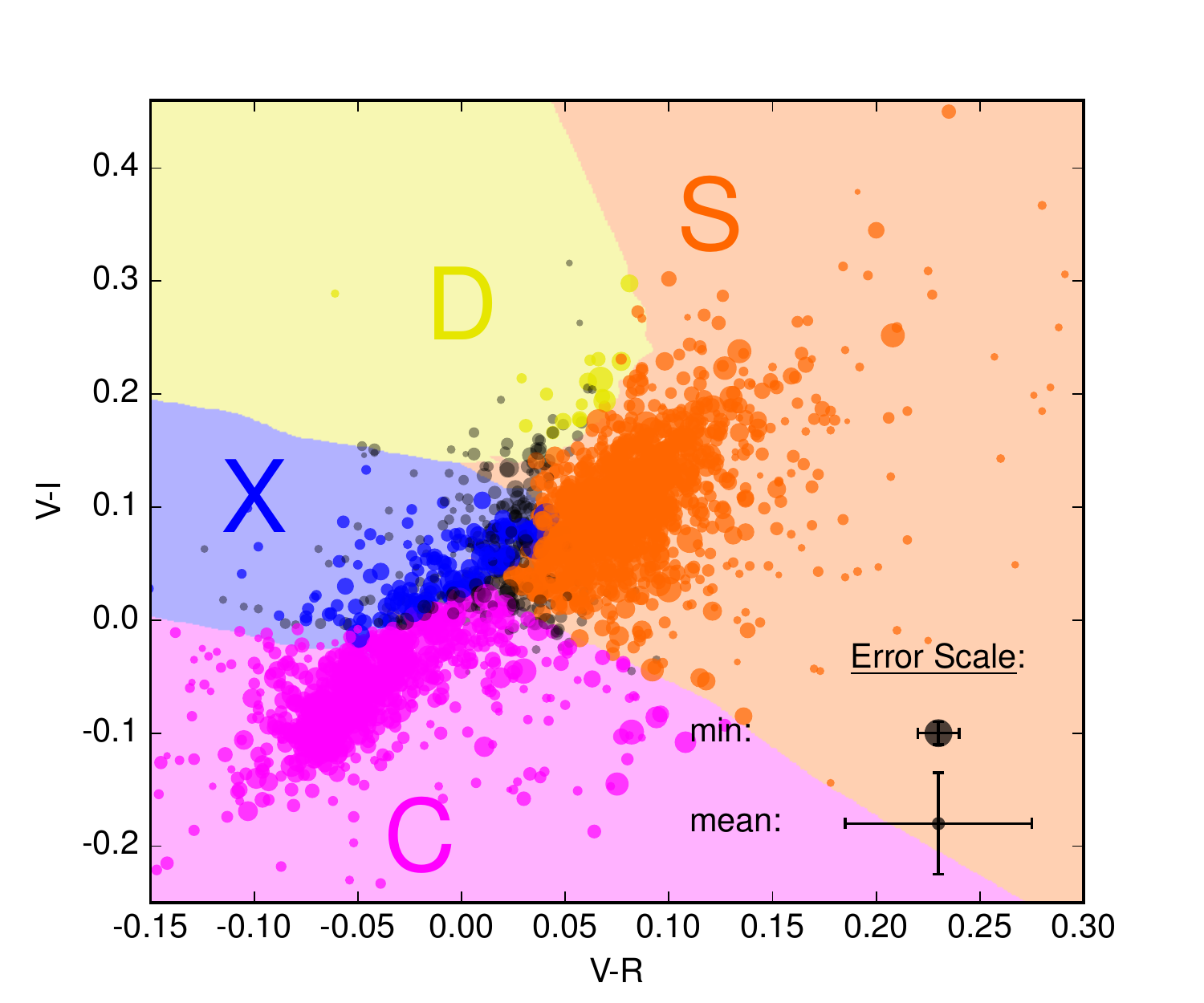}
		\caption{The measured colors of all 2276 observed MBAs  with the decision surface generated by the ML algorithm superimposed. MBAs in orange, pink, blue, and yellow were classified by the ML algorithm as S-, C-, X-, and D-type asteroids, respectively (see Section \ref{sec:color} for detail). MBAs in black remain unclassified since the highest taxonomic probability was less than 50\%. Sizes of the data points are inversely proportional to the uncertainty in calculated color (scale is shown). Taxonomic probabilities for each target are reported in a table in the Supplementary Material. (For referee, appended to manuscript as Appendix).}
		\label{all_color_color}
	\end{center}
\end{figure}

The most likely taxonomy of each target is determined by plotting the  solar-corrected \textit{V-I} versus \textit{V-R} color of each target superimposed on a decision surface in color-color space (see Figure \ref{all_color_color}). The decision surface is generated by a machine-learning (ML) algorithm trained with \textit{V-I} and \textit{V-R} colors calculated from MIT-UH-IRTF\footnote{\url{http://smass.mit.edu/minus.html}} asteroid spectra that have well defined S, C, X, or D taxonomic spectral shapes according the the Bus-DeMeo taxonomic scheme \citep{DeMeo2009}. The value of each observed target's color with respect to the decision boundaries, as well as the error in measured color, is used to calculate the probability of the target having an S, C, X, or D taxonomy. The taxonomy with the highest probability is assigned to the target and reflected in the color of the data point in Figure \ref{all_color_color} (S = orange, C = pink, X= blue, and D = yellow). Data points in black are targets with unassigned taxonomy because none of the possible taxonomies had a probability higher than 50$\%$ which we chose as a limiting criteria. This usually occurs for targets that fall within less than 1$\sigma$ in measured color to a decision boundary. For  more detail on the ML training process and method for calculating taxonomic probability see \cite{Mommert2016,Erasmus2017,Erasmus2018}. Taxonomic probabilities for each target are reported in a table in the Supplementary Material. (For referee, appended to manuscript as Appendix).

\section{Extraction of Rotational Properties}
\label{sec:rot}

Rotation periods were extracted from the light-curves by performing a Fourier analysis in the form of a least-squares spectral analysis on the combined \textit{V}, \textit{R}, and \textit{I} photometry data to generate a periodogram for each target. 
\begin{figure}[ht]
	\begin{center}
		\includegraphics[width=0.5\textwidth]{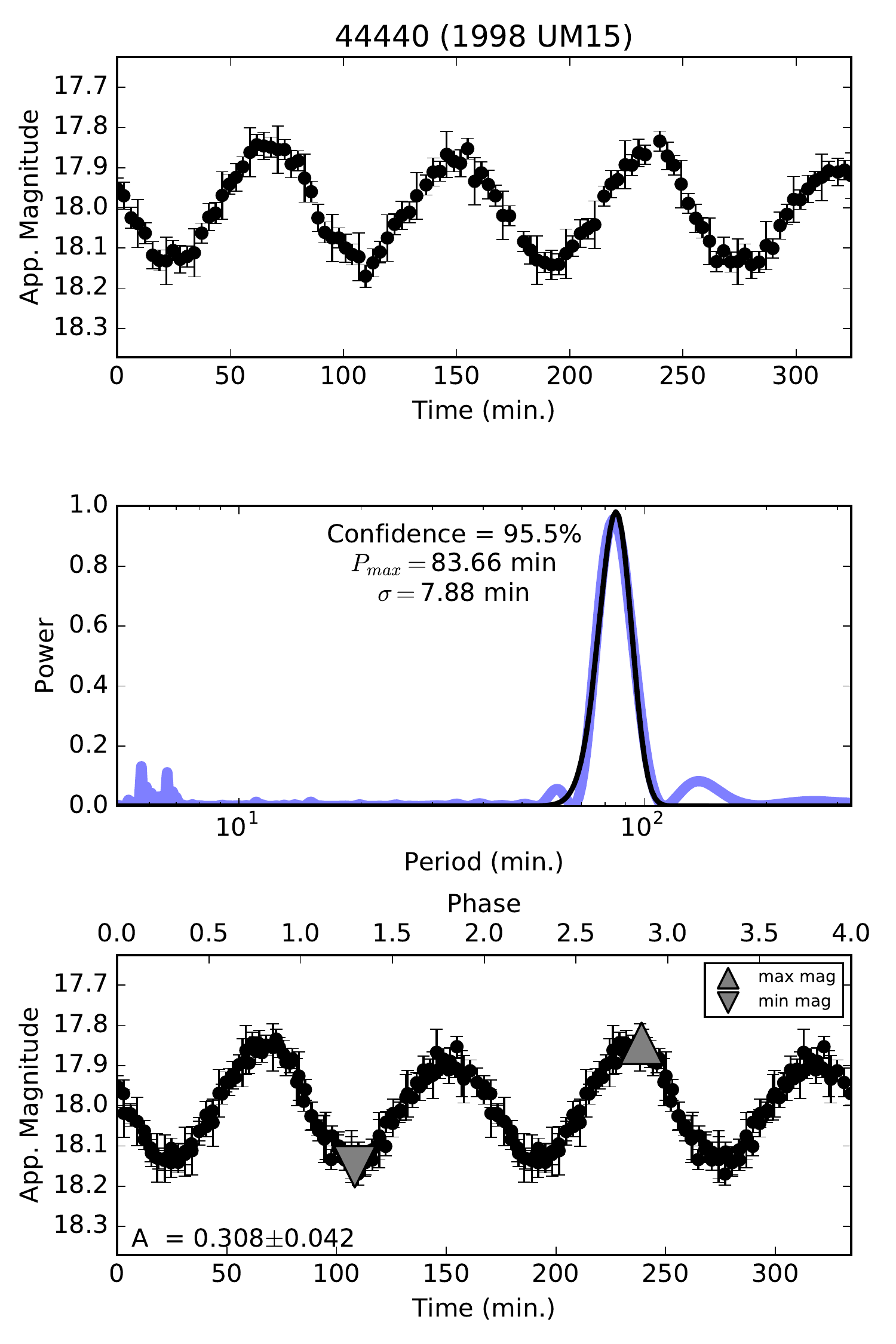}
		\caption{Example Lomb-Scargle periodogram of photometry data shown in Figure \ref{photometry_data_example}. Top: Combined \textit{V}, (adjusted)-\textit{R} and (adjusted)-\textit{I} data. Middle: Periodogram with confidence, period, and uncertainty in period shown. Bottom: Top data folded with phase=2 and plotted for phase=4. Plots for all 433 targets which had  distinct periodogram peaks and hence resolvable periods are electronically available for download. (For referee, attached as a zipped file.)}
		\label{periodogram_example}
	\end{center}
\end{figure}
An example generated Lomb-Scargle periodogram \citep{Lomb1976,Scargle1982} as a result of the analysis on the photometry data in Figure \ref{photometry_data_example} is shown in Figure \ref{periodogram_example}. Any light-curve that produced a periodogram with a distinct peak with a confidence higher than 40\% was flagged. All flagged periodograms were visually inspected and retained if deemed plausible. The uncertainty in periodicity is determined by fitting a Gaussian function to the periodogram peak and using the RMS width as the uncertainty (see superimposed black curve  on periodogram peak and $\sigma$ value in Figure \ref{periodogram_example}). Out of the 2276 targets we were able to extract periods for 433 targets. The rotation periods and light-curve amplitude for these 433 targets are recorded in the table in the Supplementary Material (For referee, appended to manuscript as Appendix). For the remaining targets we report the lower limits on the periods (i.e. observational windows) and lower limit on the light-curve amplitudes. See \cite{Erasmus2017,Erasmus2018} for more details on extraction of rotation properties.

\section{Family Determination}
\label{sec:family}
To identify an asteroid family algorithms are used to search for regions in proper orbital element space with clustering of asteroids. This is carried out using proper orbital elements rather than using the osculating orbital elements which may significantly evolve over time \citep{Knezevic2002}. The most common method used for identification of these families is the Hierarchical Clustering Method \citep{Zappala1990}.
For statistical studies it is common to simply refer to the proper orbital element boundaries for collisional families in order to insure a sufficient number of targets in each data bin. This is generally a good approximation, especially when using large data sets with sparse detections. For this work, however, we utilize and cross-correlate with data from \cite{Nesvorny2015b} obtained through The Planetary Data System\footnote{\url{https://pds.nasa.gov/}} (PDS) to associate objects from our survey data with known collisional families. We impose a cut-off of 20 objects to consider a family worthy of studying in this work and find that we have a sufficient sample for 9 collisional families. These are the Vesta,  Flora,  Nysa-Polana,  Themis,  Koronis,  Eos,  Hygiea, Massalia and Eunomia families. To identify potential interlopers within our data set we make use of the c-parameter calculated by \cite{Nesvorny2015b} for each of our observed targets. In general we find that we have extremly low levels of interloper contamination in our data set with only one potential interloper in each of the Eos, Themis, and Flora families and three potential interlopers within the Nysa-Polana family.  

\section{Results and Discussion}
\label{sec:results}

The results of our color calculations, taxonomic determinations, and rotation property extractions for all our observed targets are summarized in a table in the Supplementary Material. (For referee, appended to manuscript as Appendix). Cross-correlating our dataset with the PDS database shows that 740 of our targets are associated with a known collisional family (also indicated in the table), with 69 families being represented. Of these, 9 have sufficient ($\geq$20) targets to study the family as a sub-sample. In Section \ref{sec:compare} we compare the taxonomic distribution of family members to non-family members as a whole. In Section \ref{subsec:vesta}-\ref{subsec:eos} we compare the taxonomies for each of the 9 well-sampled collisional families we determined with the reported make-up of the family from literature. In Section \ref{subsec:parent} we define and calculate two metrics (the taxonomic purity and the degree of dispersion in observed colors) for each of the 9 well-sampled collisional families. Using these two metrics we speculate which of the 9 families potentially originated from a differentiated parent body and/or which family still has a possible undetermined nested family present.

\subsection{Comparison of Collisional Family Objects and Non-Members}
\label{sec:compare}
\begin{figure*}
	\begin{center}
		\includegraphics[width=1\textwidth]{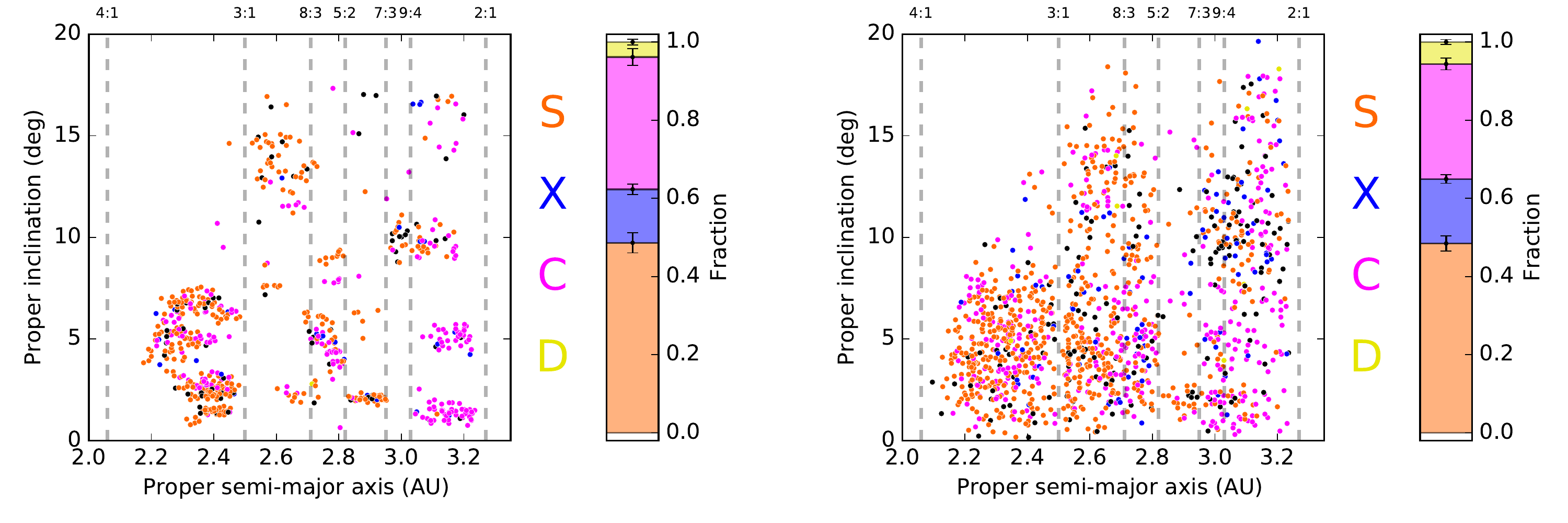}
		\caption{(Left) Plotted in proper orbital space are the 740 targets in our sample that had known collisional families after cross-correlating with data from \cite{Nesvorny2015b}. The color of each data point indicates the taxonomy we determine (see Section \ref{sec:color}) with orange,  blue,  pink  and  yellow for S-, X- and C- and D-type asteroids,  respectively. In general, the plot shows clustering of similar taxonomic types in proper orbital space which is in agreement with the notion that there should be homogeneity in composition within families. (Right) Plotted in proper orbital space are the remaining targets in our sample that had no known collisional families after cross-correlating with data from \cite{Nesvorny2015b}. (Left and Right) Adjacent to the main plots, the taxonomic distributions of the family and non-family populations are shown. Within uncertainties the two populations have identical distributions.}
		\label{compare}
	\end{center}
\end{figure*}

We compare the derived taxonomies for objects linked to known collisional families and those remaining objects which are not. If the taxonomic proportions of these two populations were different this could imply that certain asteroid compositions are more frequently disrupted by catastrophic collisions leading to the formation of asteroid families. In this case we find that the overall proportions of taxonomies in the two populations are identical to within uncertainties. This is presented in Figure~\ref{compare}.

\subsection{Individual Family Analysis}
\label{sec:indiv}

\subsubsection{Vesta}
\label{subsec:vesta}
The Vesta family is one of the most widely-studied main belt sub-populations. Earth-based spectrometric observations have shown a distinct spectral signature that is similar to the spectral shape of S-type asteroids but with enhanced absorption features in both the 1$\mu$m and 2$\mu$m regions. 

\begin{figure}
	\begin{center}
		\includegraphics[width=0.5\textwidth]{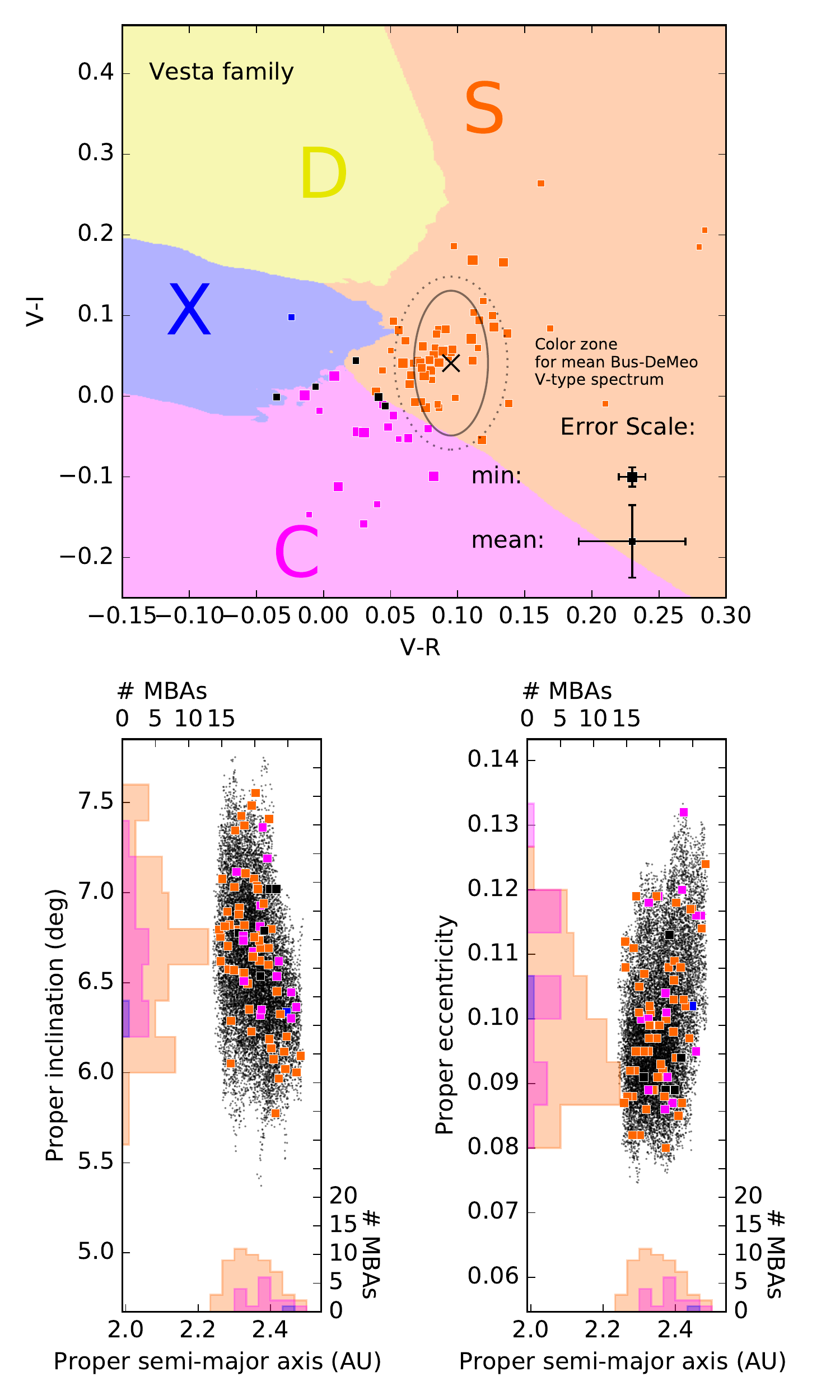}
		\caption{(Top) The measured colors of all 76 observed Vesta targets  with the taxonomic decision surface generated by the ML algorithm superimposed (see Section \ref{sec:color} for detail). The oval with the cross in the centre indicates the color and limits calculated from the mean Bus-DeMeo V-type spectrum. The dashed oval reflects the extent of the limits when incorporating our median observed photometric uncertainty. (Bottom) Proper orbital elements of all 76 observed Vesta targets with histograms indicating the taxonomic dependence on orbital parameters. Family members from \cite{Nesvorny2015b} are plotted in small data points in the background.}
		\label{Vesta_color_color}
	\end{center}
\end{figure}

\begin{figure}
	\begin{center}
	\includegraphics[width=0.5\textwidth]{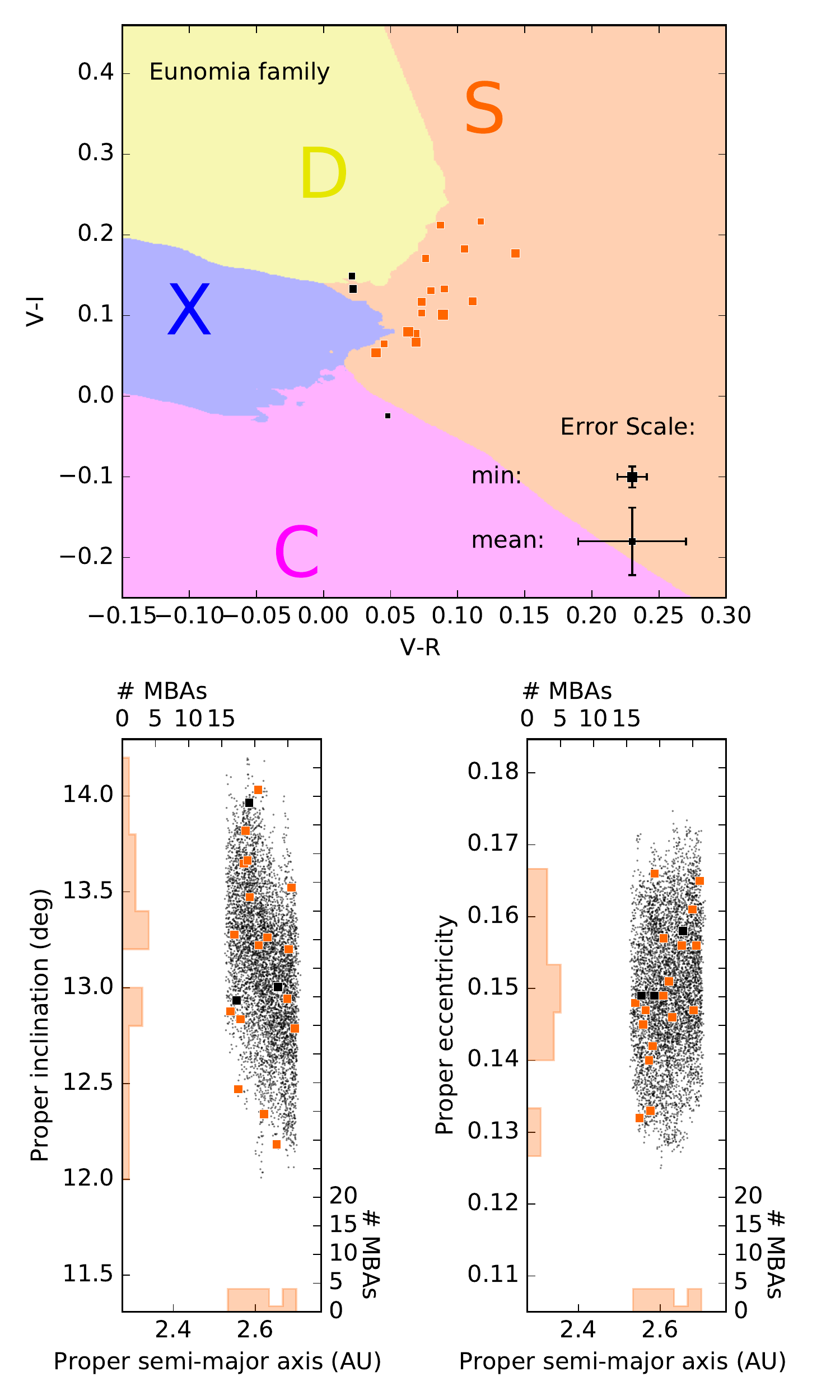}
	\caption{(Top) The measured colors of all 20 observed Eunomia targets  with the taxonomic decision surface generated by the ML algorithm superimposed (see Section \ref{sec:color} for detail). (Bottom) Proper orbital elements of all 20 observed Eunomia targets with histograms indicating the taxonomic dependence on orbital parameters. Family members from \cite{Nesvorny2015b} are plotted in small data points in the background.}
	\label{Hygiea_color_color}
	\label{Eunomia_color_color}
	\end{center}
\end{figure}

In Figure \ref{Vesta_color_color} we plot the colors of the 76 Vesta family targets observed in this study superimposed on the decision surface generated by the ML algorithm (see section \ref{sec:color}). Unfortunately  the possibility of a V-type classification was not included in the ML classification method as there are insufficient visible spectra of V-type asteroids in the MIT-UH-IRTF database which is the source of the training data. The consequence of this is that the ML method mostly classifies V-type asteroids as S-type asteroids because of the similar spectral shape between the two spectra. However, an oval is included in Figure \ref{Vesta_color_color} indicating where one would expect the visible colors of a V-type asteroid. This was achieved by convolving the \textit{V}-, \textit{R}-, and \textit{I}-filter responses with the mean Bus-DeMeo V-type spectrum \citep{DeMeo2009}, with the associated upper- and lower-limits, to calculate the mean, minimum, and maximum \textit{V-R} and \textit{V-I} colors expected from this spectrum. As expected the oval region occurs within the S-type zone but at a lower \textit{V-I} color than the mean S-type color (see Figure \ref{all_color_color}) due to the stronger 1$\mu$m absorption feature of V-type spectra. Most of the observed Vesta family targets cluster in the vicinity of the expected V-type region but in comparison to the other 8 families (Figure \ref{Flora_color_color}-\ref{Eos_color_color}) we observe a larger spread in colors. In addition we observe a significant number of our targets (with small uncertainties in measured color) falling within the C-type zone in color-color space. Our result therefore supports the idea behind a differentiated parented body with multiple taxonomies possible in the daughter bodies \citep{Russell2012}.

\subsubsection{Flora}
\label{subsec:flora}

The Flora family, located in the inner main belt, has been well-established as an S-type family \citep{Florczak1998}. Situated in orbital space close to the $\nu_6$ resonance it has been identified as a possible source for the near-Earth asteroid (NEA) population \citep{Bottke2000a}. Similarities in spectral properties between the NEA population and the Flora family have also been been confirmed \citep{Vernazza2008}.

\begin{figure}
	\begin{center}
		\includegraphics[width=0.5\textwidth]{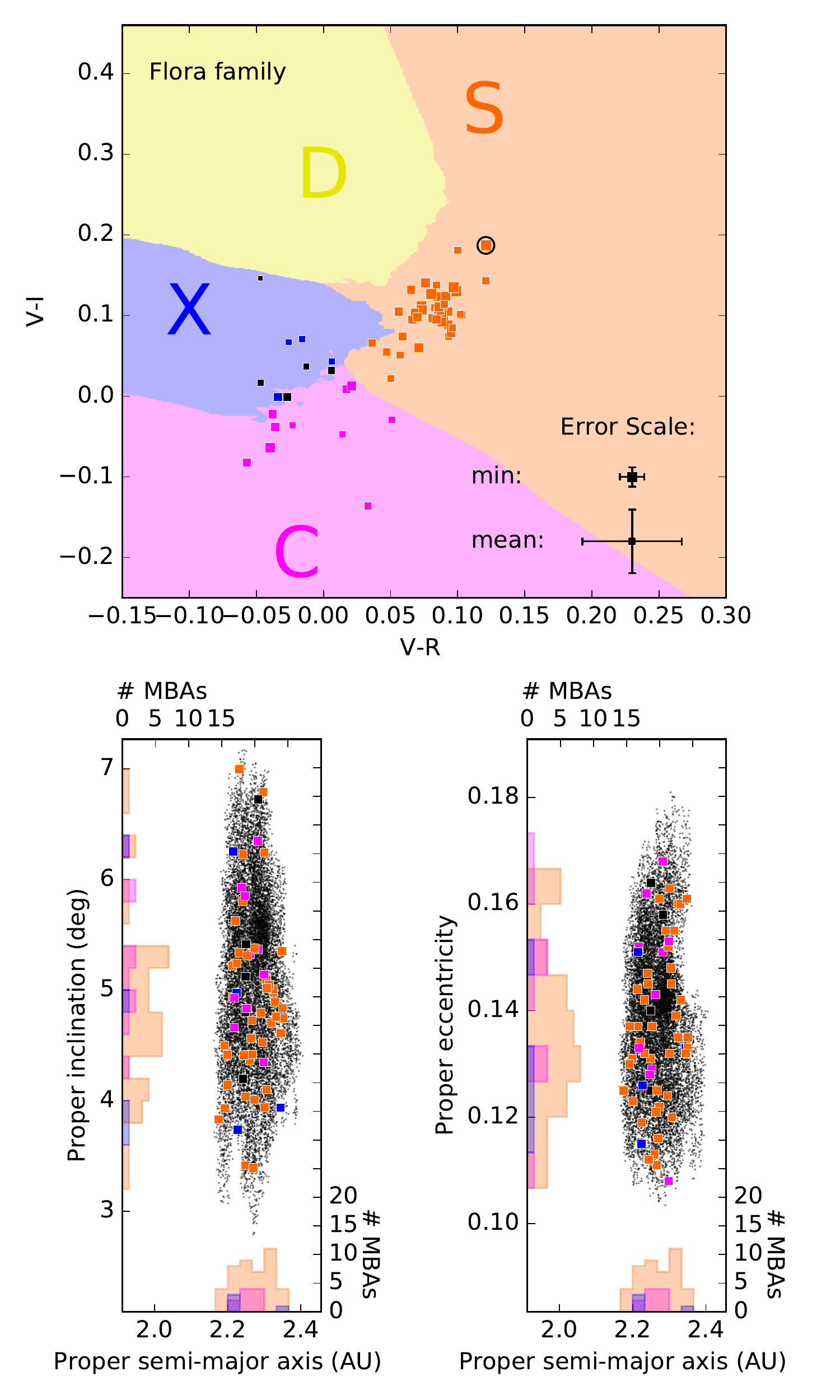}
		\caption{(Top) The measured colors of all 62 observed Flora targets with the taxonomic decision surface generated by the ML algorithm superimposed (see Section \ref{sec:color} for detail). The circled data point is a suspected interloper because the c-parameter (see \cite{Nesvorny2015} for definition and detail) for this target has an absolute value larger than 1. (Bottom) Proper orbital elements of all 62 observed Flora targets with histograms indicating the taxonomic dependence on orbital parameters. Family members from \cite{Nesvorny2015b} are plotted in small data points in the background.}
		\label{Flora_color_color}
	\end{center}
\end{figure}

The colors plotted in Figure \ref{Flora_color_color} of the 62 observed Flora family members show a majority in S-type classification and are therefore in agreement with Flora family literature. However,  there are a few targets with C-type colors observed. We find the compositional break-down of the Flora family to be within uncertainties identical to the compositional break-down of a separate study of 39 NEAs using the same ML classification technique used in this study \citep{Erasmus2017}. Our taxonomic distribution results here therefore support previously published works suggesting the Flora family as a strong feeder for the NEA population. The large number of C-type asteroids in our sample could suggest the presence of a still undetermined nested family or evidence that Flora originated from a differentiated parent body (see Section \ref{subsec:parent}).

\subsubsection{Massalia}
\label{subsec:massalia}
Located in the inner main belt, the relatively young ($180 \pm 50$ Myr; \cite{Spoto2015}) Massalia family is distinguishable in orbital space from other families by its low ($\approx$ $1.5^{\circ}$) orbital inclination. It is thought that the Massalia family can be a source region for NEAs with objects being ejected from the main belt through the 3:1 mean motion resonance \citep{Milani2014}.

The derived taxonomies for the Massalia family are in agreement with the expected S-type nature \citep{Masiero2015} of the family to within uncertainties.

\begin{figure}
	\begin{center}
		\includegraphics[width=0.5\textwidth]{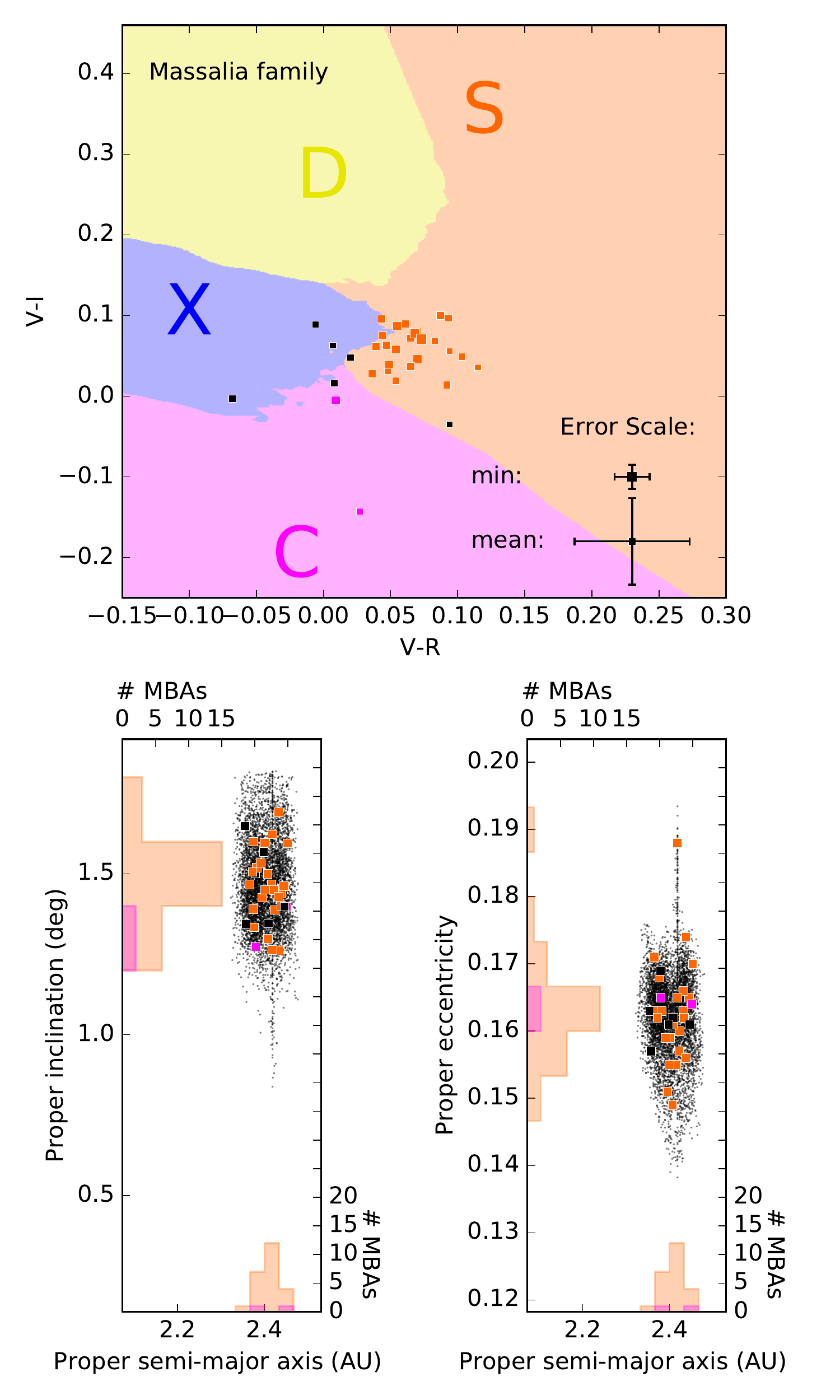}
		\caption{(Top) The measured colors of all 32 observed Massalia targets  with the taxonomic decision surface generated by the ML algorithm superimposed (see Section \ref{sec:color} for detail). (Bottom) Proper orbital elements of all 32 observed Massalia targets with histograms indicating the taxonomic dependence on orbital parameters. Family members from \cite{Nesvorny2015b} are plotted in small data points in the background.}
		\label{Massalia_color_color}
	\end{center}
\end{figure}

\subsubsection{Nysa-Polana}
\label{subsec:nysa-polana}
The Nysa-Polana family is no longer regarded as a single entity but rather as two subgroups with similar proper orbital properties but with unambiguously distinct albedo and spectral signatures \citep{Cellino2001}. The eponym of the family, (44) Nysa, has subsequently also been excluded as a member of either subgroup based on a combination of its differing albedo and reflective spectrum to the albedo and spectra of the two identified subgroups \citep{Masiero2015}. That said, the ``Nysa'' subgroup is associated with the S-type taxonomic class while the ``Polana'' subgroup has shown B-type spectral signatures.

\begin{figure}
	\begin{center}
		\includegraphics[width=0.5\textwidth]{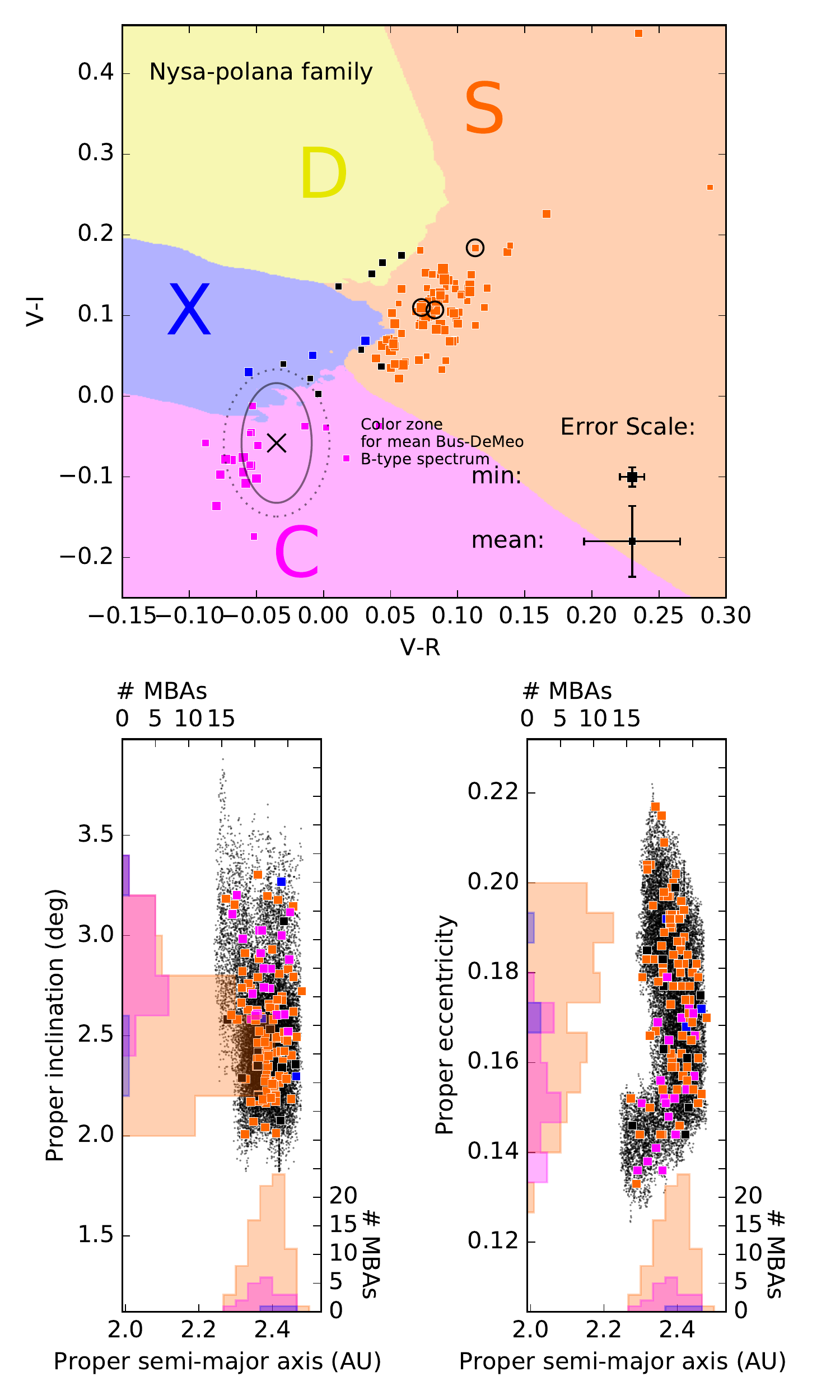}
		\caption{(Top) The measured colors of all 118 observed Nysa-Polana targets  with the taxonomic decision surface generated by the ML algorithm superimposed (see Section \ref{sec:color} for detail). The three circled data points are suspected interlopers because the c-parameters (see \cite{Nesvorny2015} for definition and detail) for these targets have an absolute value larger than 1. The oval with the cross in the centre indicates the color and limits calculated from the mean Bus-DeMeo B-type spectrum. The dashed oval reflects the extent of the limits when incorporating our median observed photometric uncertainty. (Bottom) Proper orbital elements of all 118 observed Nysa-Polana targets with histograms indicating the taxonomic dependence on orbital parameters. Family members from \cite{Nesvorny2015b} are plotted in small data points in the background.}
		\label{Nysa-Polana_color_color}
	\end{center}
\end{figure}

Just as our ML classification method could not distinguish between the V-type taxonomic class and the S-type taxonomic class (see Section \ref{subsec:vesta}), the method is also unable to distinguish between B-type and C-type asteroids. As was done for expected V-type colors, Figure \ref{Nysa-Polana_color_color} shows the expected region for B-type asteroids in color-color space (see black oval in plot). Figure \ref{Nysa-Polana_color_color} also includes the plotted colors for the 118 observed Nysa-Polana family members. The plotted colors clearly reveal the two subgroups, one group centred around the mean S-type visible colors and another tightly clustered group in close proximity to the expected B-type visible colors.

\subsubsection{Eunomia}
\label{subsec:eunomia}

Eunomia is a primarily S-type family found in the intermediate main belt thought to have formed $1.5 \pm 0.5$ Gyr ago \citep{Spoto2015}. \cite{Lazzaro1999} identified the presence of objects in the same proper orbital element space as the Eunomia family showing featureless spectra indicative of C-type asteroids. These objects were suggested to either be interlopers or evidence for a differentiated parent body. \cite{Milani2014} suggest that the Eunomia family consists of further sub-families caused by cratering events within the lifetime of the family as a whole. From our limited sample we do not identify any clear indication of a large number of outliers from the expected S-type population. The derived taxonomies suggest that to within uncertainties all our observed objects can be considered S-type asteroids.

\subsubsection{Hygiea}
\label{subsec:Hygiea}

The Hygiea family is a primarily C-type family found in the outer main asteroid belt, although B-type asteroids are also thought to be present \citep{Carruba2013}. The parent body of this family (10) Hygiea is the fourth largest asteroid (by mass) in the main belt. As expected our taxonomic determination sits within the boundary for C-types. Some objects in the sample have been identified as X-types, however, the uncertainties on this determination are such that we are unable to conclude with any confidence that these bodies are true outliers. Due to the dearth of training data available for B-type asteroids we are unable to directly determine whether objects are truly B-types or C-types; we have plotted the boundaries in this color-color plot in which B-type asteroids would fall. We find that all of the objects in our sample fall within the expected C-type boundary to within uncertainties, with several falling into the region described by the B-type estimate although we cannot conclusively classify these objects with that degree of detail with the current data available for our methodology.

\begin{figure}
	\begin{center}
	\includegraphics[width=0.5\textwidth]{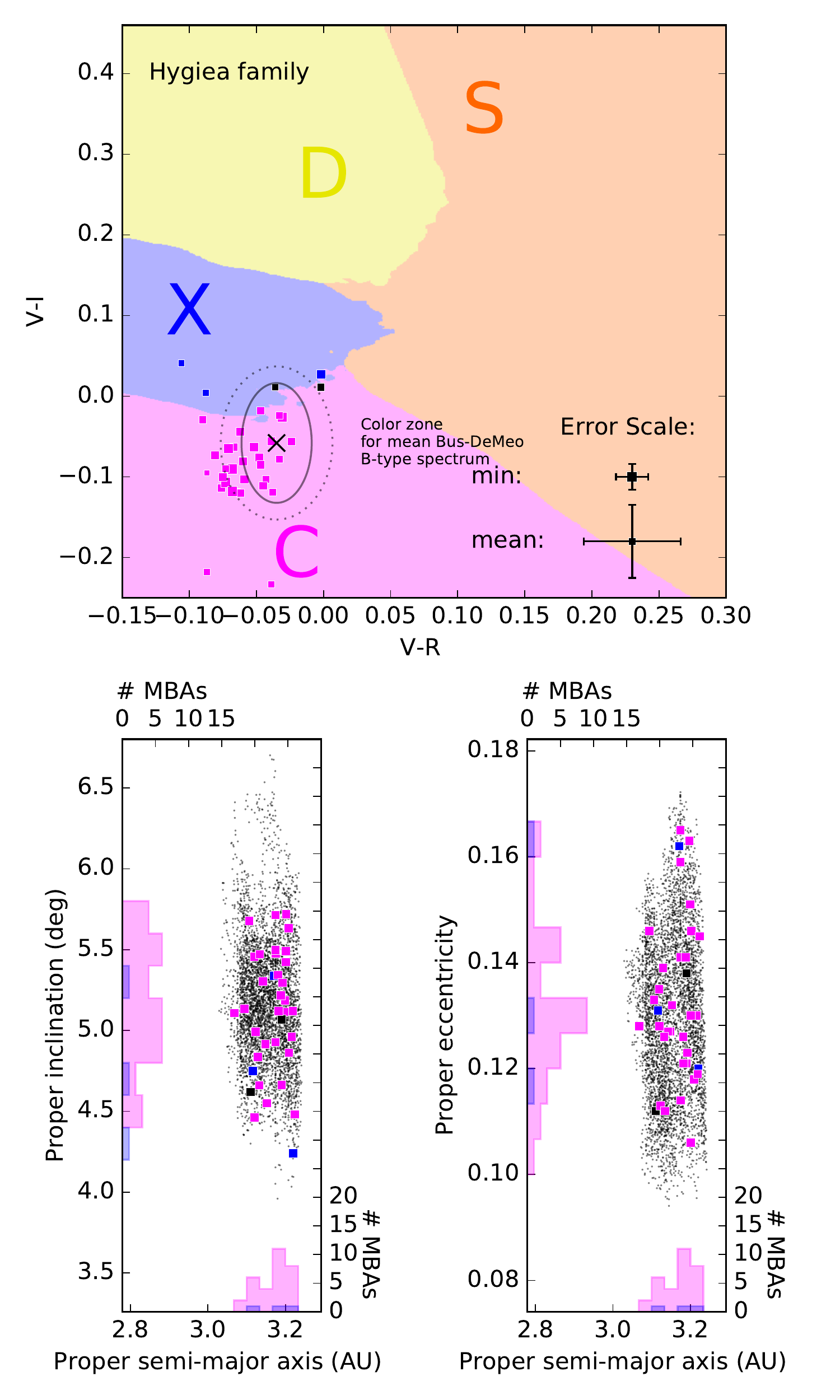}
	\caption{(Top) The measured colors of all 36 observed Hygiea targets  with the taxonomic decision surface generated by the ML algorithm superimposed (see Section \ref{sec:color} for detail). The oval with the cross in the centre indicates the color and limits calculated from the mean Bus-DeMeo B-type spectrum. The dashed oval reflects the extent of the limits when incorporating our median observed photometric uncertainty. (Bottom) Proper orbital elements of all 36 observed Hygiea targets with histograms indicating the taxonomic dependence on orbital parameters. Family members from \cite{Nesvorny2015b} are plotted in small data points in the background.}
	\label{Hygiea_color_color}
	\end{center}
\end{figure}

\subsubsection{Themis}
\label{subsec:themis}

The Themis family, like most families located on the outer edge of the main-belt, has been observed to be comprised of mostly C-type members \citep{Ziffer2011}. The discovery of cometary activity on at least two family members \citep{Hsieh2006} as well as direct evidence of water-ice on (24) Themis itself by \cite{Rivkin2010} suggests that this family could be a significant reservoir of water-ice in the solar system.

\begin{figure}
	\begin{center}
		\includegraphics[width=0.5\textwidth]{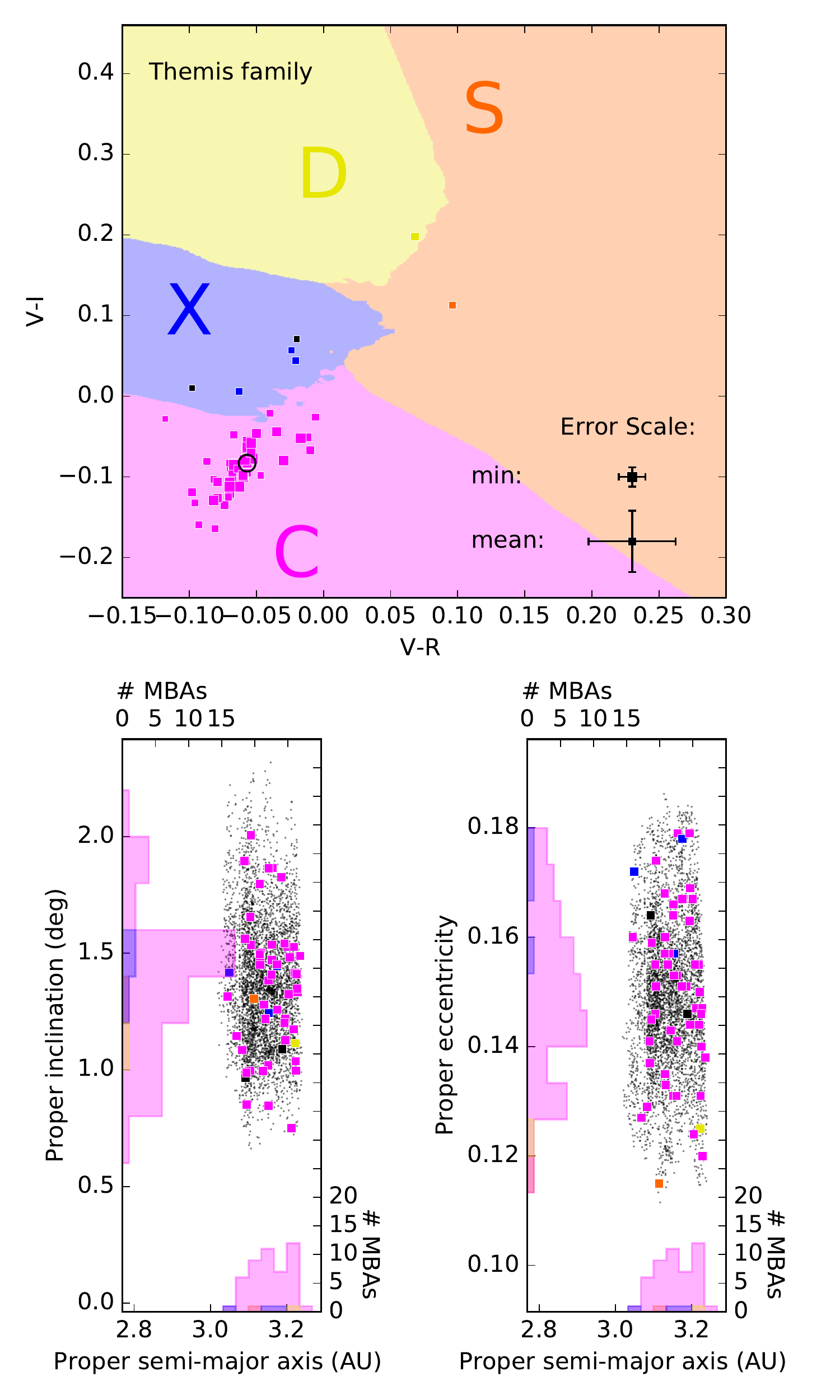}
		\caption{(Top) The measured colors of all 54 observed Themis targets  with the taxonomic decision surface generated by the ML algorithm superimposed (see Section \ref{sec:color} for detail). The circled data point is a suspected interloper because the c-parameter (see \cite{Nesvorny2015} for definition and detail) for this target has an absolute value larger than 1. (Bottom) Proper orbital elements of all 54 observed Themis targets with histograms indicating the taxonomic dependence on orbital parameters. Family members from \cite{Nesvorny2015b} are plotted in small data points in the background.}
		\label{Themis_color_color}
	\end{center}
\end{figure}

The visible colors of the 54 Themis family targets observed in this study imply that the targets are almost exclusively C-type in taxonomy (see Figure \ref{Themis_color_color}), which is in agreement with previous studies and observations. We find that, as expected, most objects are defined as being C-type asteroids with a high degree of confidence. Several objects are classified as X-types but the uncertainties on these measurements are such that they cross the boundary between X- and C-type. We identify two objects with taxonomic determinations significantly differing from that of the main family which we suspect to be interlopers in the Themis family. They are 9646 (1995 BV) and 187245 (2005 SV190) which were determined to be D-type and S-type respectively.

\subsubsection{Koronis}
\label{subsec:koronis}

\begin{figure}
	\begin{center}
	\includegraphics[width=0.5\textwidth]{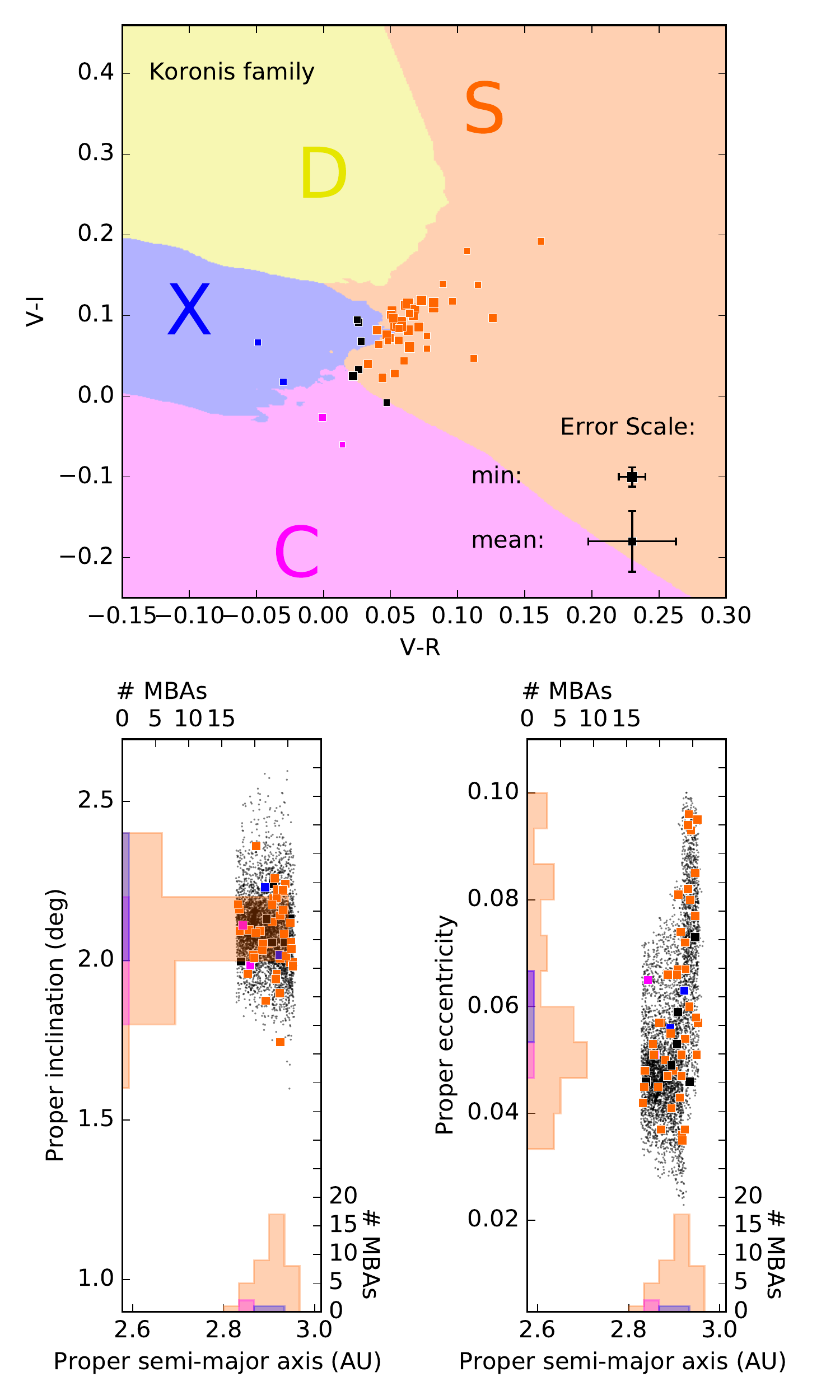}
	\caption{(Top) The measured colors of all 50 observed Koronis targets  with the taxonomic decision surface generated by the ML algorithm superimposed (see Section \ref{sec:color} for detail). (Bottom) Proper orbital elements of all 50 observed Koronis targets with histograms indicating the taxonomic dependence on orbital parameters. Family members from \cite{Nesvorny2015b} are plotted in small data points in the background.}
	\label{Koronis_color_color}
	\end{center}
\end{figure}

The Koronis family is situated in the outer main belt and, unusually for its semi-major axis location, consists primarily of S-type asteroids. We find that most objects belonging to this family in our sample are determined to be S-type objects (see Figure \ref{Koronis_color_color}).

\subsubsection{Eos}
\label{subsec:eos}

\begin{figure}
	\begin{center}
		\includegraphics[width=0.5\textwidth]{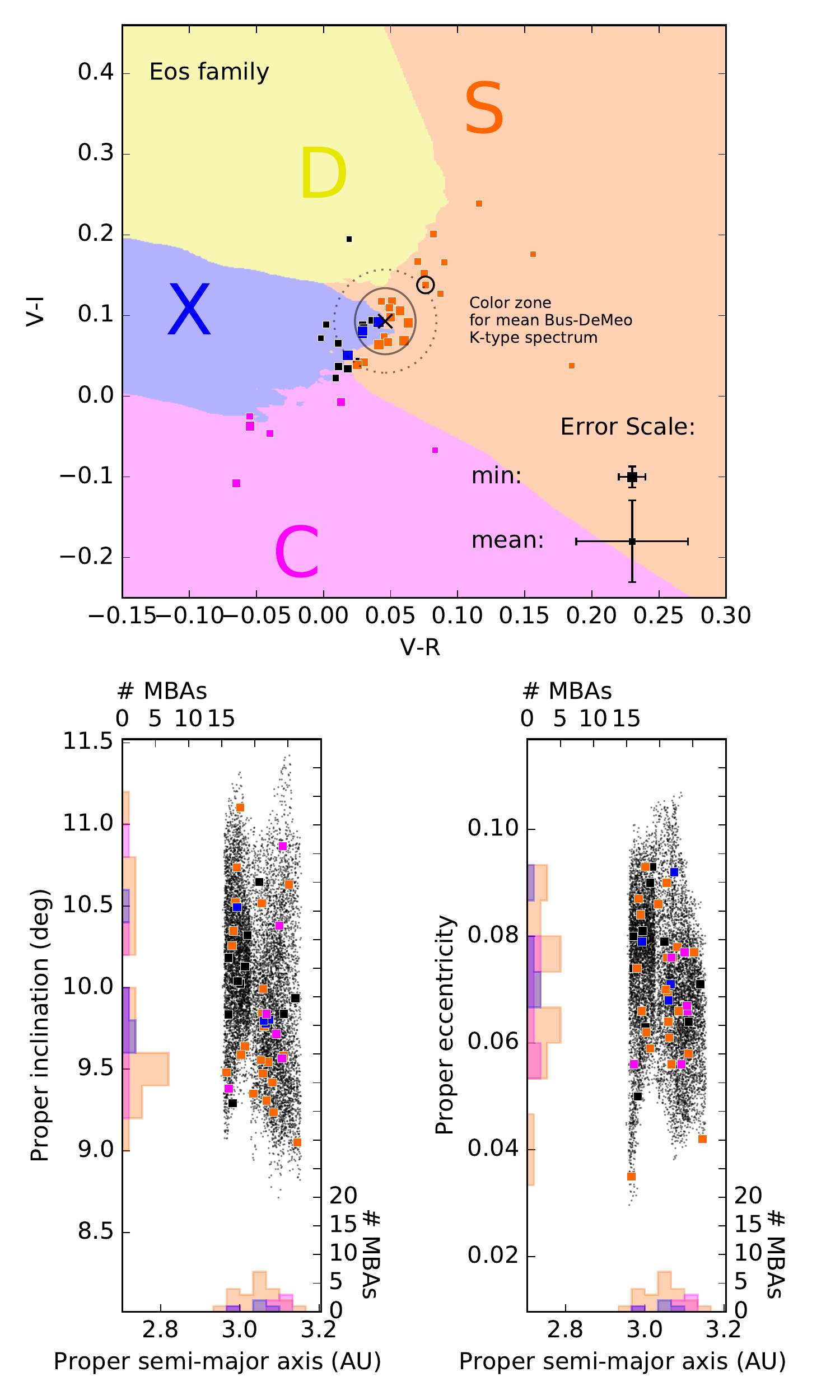}
		\caption{(Top) The measured colors of all 43 observed Eos targets  with the taxonomic decision surface generated by the ML algorithm superimposed (see Section \ref{sec:color} for detail). The circled data point is a suspected interloper because the c-parameter (see \cite{Nesvorny2015} for definition and detail) for this target has an absolute value larger than 1. The oval with the cross in the centre indicates the color and limits calculated from the mean Bus-DeMeo K-type spectrum. The dashed oval reflects the extent of the limits when incorporating our median observed photometric uncertainty. (Bottom) Proper orbital elements of all 43 observed Eos targets with histograms indicating the taxonomic dependence on orbital parameters. Family members from \cite{Nesvorny2015b} are plotted in small data points in the background.}
		\label{Eos_color_color}
	\end{center}
\end{figure}

The Eos family is comprised of the relatively rare K-type asteroid taxonomy associated with CV and CO meteorite spectra. The K-type spectral shape is similar to that of S-type spectra but with a slightly smaller red-slope towards infra-red wavelengths and the absence of the 2$\mu$m absorption feature that is present in S-type spectra. Therefore it is difficult to distinguish K-type spectra from S-type spectra in the visible where there is little difference between the two. As a result, the 43 Eos family targets have determined colors that classify them as mostly S-type asteroids by our ML method as shown in Figure \ref{Eos_color_color}. Many of the targets with low uncertainty in their calculated color fall within the expected K-type region (see black oval, and \ref{subsec:vesta} for method to determining perimeter) which suggests that the observations agree with the K-type taxonomy claimed by current literature. The lack of training data available for K-type asteroids means that this conclusion must be considered speculative.

\subsection{Identifying Differentiated Parent Bodies and Nested Families}
\label{subsec:parent}

In order to identify which of our well-sampled families possibly originate from a differentiated parent body, we define and calculate two metrics using the observed colors of each of our targets. The first is the purity in our observed taxonomy for a given family which is defined as the highest taxonomic fraction based on the summation of the probabilities for each respective class type (see table in Supplementary material for probabilities). The second metric quantifies how dispersed the observed colors are by calculating the median difference in color (weighted by error in measured color) from the median color of all the observed targets of the family. These two metrics potentially single out families with differentiated parents since a non-differentiated parent would likely result in low variations in observed colors and also a high purity in measured taxonomy of daughter bodies.

For this analysis we treat the Nysa-Polana family as two separate families (``Nysa'' and ``Polana'') since this is now commonly accepted as being two separate families  with differing taxonomy (the two different taxonomic groups are observed in this study as well, see Figure \ref{Nysa-Polana_color_color}). However, we also include a datapoint for the combined Nysa-Polana family in order to get an idea of what the two metric values would be for a bi-modal (nested) family like Nysa-Polana. Finally, we also exclude the Eos family because the expected (and measured, see Figure \ref{Eos_color_color}) colors of K-type asteroids fall on top of the decision boundary between S- and X-type colors. Since we do not include K-type as a possible classification in our analysis the the purity metric makes little sense in this case. We do not encounter this purity determination problem for the Vesta family (V-type) or the ``Polana'' family (B-type) as both of these colors are encapsulated in the S- and C-type zones of the color-color plot respectively and do not fall close to a decision boundary (see Figure \ref{Vesta_color_color} and \ref{Nysa-Polana_color_color}). We therefore use the S-type fraction as a purity metric for the Vesta family and the C-type fraction as the purity metric for the ``Polana'' family.

\begin{figure}
	\begin{center}
		\includegraphics[width=0.5\textwidth]{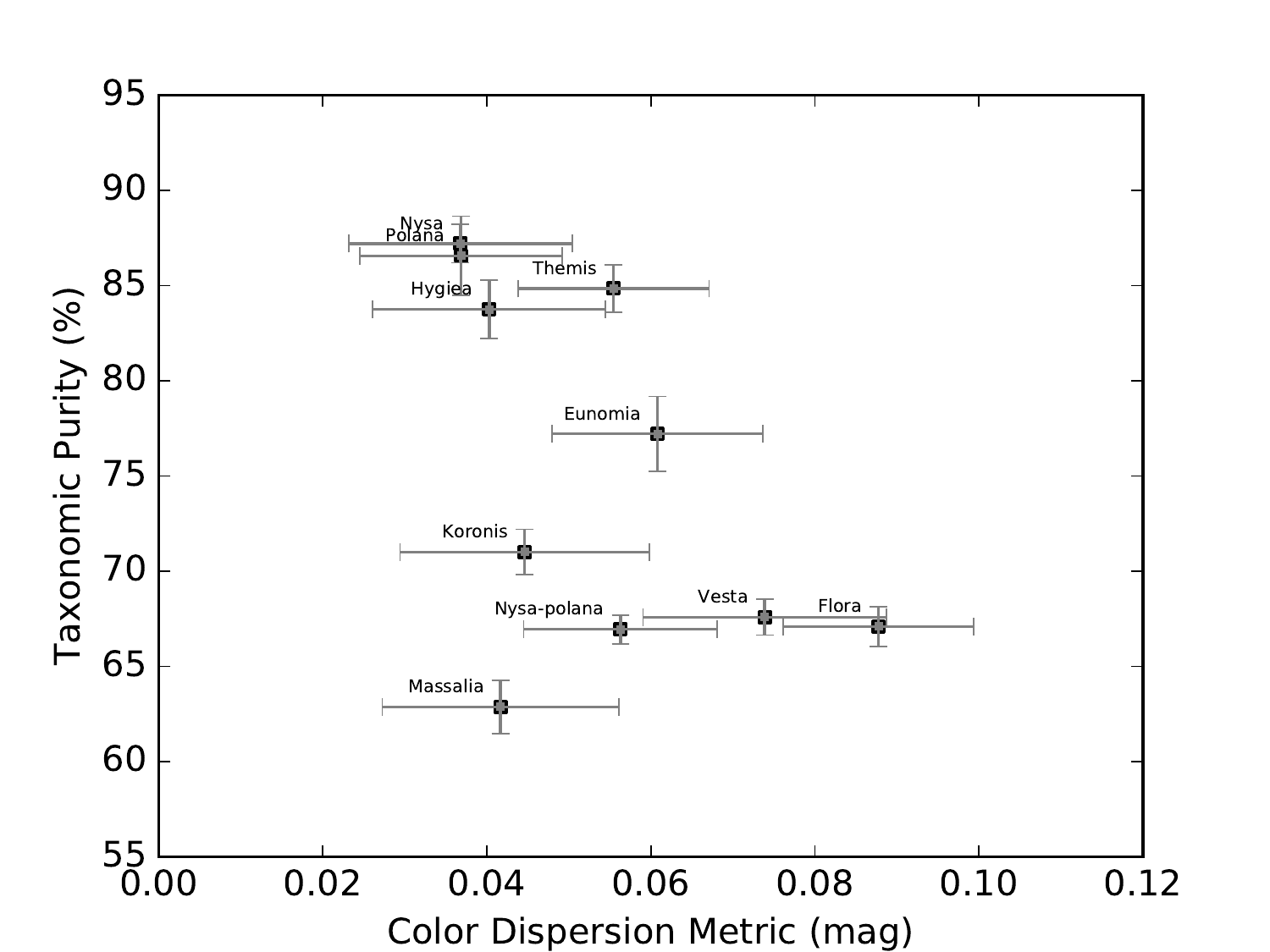}
		\caption{Plotted is the taxonomic purity and the degree of dispersion in observed colors for each of the 9 well-sampled collisional families. See Section \ref{subsec:parent} for definition and discussion of these two metrics.}
		\label{color_clustering_metric}
	\end{center}
\end{figure}

In Figure \ref{color_clustering_metric} we plot the results of our analysis and as expected the Vesta family which has been confirmed by the Dawn mission to originate from a differentiated parent body \citep{DeSanctis2012} has a low purity in observed taxonomy and high variations in observed colors compared to most of the other families we plot. The bi-modal Nysa-Polana family's metrics also shows low purity in observed taxonomy and high variations in observed colors but when separated into the ``Nysa'' and ``Polana'' family show very high purity and low variation in colors. The remaining families, with the exception of Massalia and Flora, have both a higher taxonomic purity and a lower color dispersion than the Vesta and Nysa-Polana family suggesting that none of these originate from differentiated parent body or have a nested family with a differing taxonomy present. 

Our Massalia family targets have the highest mean error in determined color of the 9 families we investigate (compare error bar scale in Figure \ref{Massalia_color_color} to that of the other families) and probably the reason for the low purity metric we determine. Therefore the Massalia family's purity metric is the least reliable of all the families and probably not dependable enough to make any concrete conclusions regarding its parent body. 

The Flora family has very similar metric values to the Vesta family which could be an indication that this family also originated from a differentiated parent body. A differentiated parent body for the Flora family has been suggested in the past \citep{Gaffey1984}. Using Nysa-Polana, which shows evidence of a bimodal  histogram (one C-type peak and one S-type peak, see Figure \ref{Nysa-Polana_color_color}) in both eccentricity and inclination as a yardstick for a nested family, and comparing that to the histograms of Flora (see Figure \ref{Flora_color_color}) which show no evidence of any bimodal behaviour further supports the idea that Flora's observed mixed taxonomy is due to a differentiated parent body rather than a nested family.

\section{Conclusion}
Multi-band photometry for 2276 MBAs has been presented. For each target we determine the probabilistic taxonomy using the measured \textit{V-R} and \textit{V-I} colors in combination with a machine-learning generated decision surface in color-color space. We investigate in detail our determined taxonomy of all targets of 9 families which had 20 or more members present in our data set. In general we find excellent agreement with the observed taxonomy of these families to the expected taxonomy from previous studies. 

Using the observed colors for each target we define and calculate two metrics of the 9 well-sampled collisional families and use these two metrics to speculate which of these families potentially originated from a differentiated parent body and/or is a family with a possible undetermined nested family. Our results show that the Flora family has very similar metric values to the Vesta family which could be an indication that, like the Vesta family, the Flora family possibly also originated from a differentiated parent body.

Finally, our photometry data was sufficient to extract reliable rotation periods for 433 of our targets and we observe no obvious correlation between rotation properties and family membership. 

\section*{Acknowledgements}

This research has made use of the KMTNet system operated by the Korea Astronomy and Space Science Institute (KASI) and the data were obtained by observations made at the South African Astronomical Observatory (SAAO). This work is partially supported by the South African National Research Foundation (NRF). This work is supported in part by the National Aeronautics and Space Administration (NASA) under grant No. NNX15AE90G issued through the SSO Near-Earth Object Observations Program and in part by a grant from NASA's Office of the Chief Technologist.

\bibliography{N_Erasmus_Asteroid_Families_2019.bib}

\begin{thebibliography}{}
\expandafter\ifx\csname natexlab\endcsname\relax\def\natexlab#1{#1}\fi

\bibitem[{Berthier {et~al.}(2006)Berthier, Vachier, \& Thuillot}]{Berthier2006}
Berthier, J., Vachier, F., \& Thuillot, W. 2006, 351, 367

\bibitem[{Binney \& Merrifield(1998)}]{Binney1998}
Binney, J., \& Merrifield, M. 1998, {Galactic Astronomy}

\bibitem[{Binzel \& Xu(1993)}]{Binzel1993}
Binzel, R.~P., \& Xu, S. 1993, Science, 260, 186

\bibitem[{Bottke {et~al.}(2000)Bottke, Rubincam, \& Burns}]{Bottke2000a}
Bottke, W.~F., Rubincam, D.~P., \& Burns, J.~A. 2000, Icarus, 145, 301

\bibitem[{Brunetto \& Strazzulla(2005)}]{Brunetto2005}
Brunetto, R., \& Strazzulla, G. 2005, Icarus, 179, 265

\bibitem[{Bus(1999)}]{Bus1999}
Bus, S. 1999, PhD thesis, MASSACHUSETTS INSTITUTE OF TECHNOLOGY

\bibitem[{Carruba(2013)}]{Carruba2013}
Carruba, V. 2013, MNRAS, 431, 3557

\bibitem[{Cellino {et~al.}(2001)Cellino, Zappal{\`{a}}, Doressoundiram, {Di
  Martino}, Bendjoya, Dotto, \& Migliorini}]{Cellino2001}
Cellino, A., Zappal{\`{a}}, V., Doressoundiram, A., {et~al.} 2001, Icarus, 152,
  225

\bibitem[{{De Sanctis} {et~al.}(2012){De Sanctis}, Ammannito, Capria, Tosi,
  Capaccioni, Zambon, Carraro, Fonte, Frigeri, Jaumann, Magni, Marchi, Mccord,
  Mcfadden, McSween, Mittlefehldt, Nathues, Palomba, {M Pieters}, \&
  Turrini}]{DeSanctis2012}
{De Sanctis}, M.~C., Ammannito, E., Capria, M., {et~al.} 2012, Science (New
  York, N.Y.), 336, 697

\bibitem[{DeMeo {et~al.}(2009)DeMeo, Binzel, Slivan, \& Bus}]{DeMeo2009}
DeMeo, F.~E., Binzel, R.~P., Slivan, S.~M., \& Bus, S.~J. 2009, Icarus, 202,
  160

\bibitem[{Dermott {et~al.}(2018)Dermott, {A. Christou}, Li, {J. J. Kehoe}, \&
  {Malcolm Robinson}}]{Dermott2018}
Dermott, S., {A. Christou}, A., Li, D., {J. J. Kehoe}, T., \& {Malcolm
  Robinson}, J. 2018, Nature Astronomy, 2, doi:10.1038/s41550-018-0482-4

\bibitem[{Erasmus {et~al.}(2018)Erasmus, Mcneill, \& Mommert}]{Erasmus2018}
Erasmus, N., Mcneill, A., \& Mommert, M. 2018, The Astrophysical Journal
  Supplement Series, 237, 19

\bibitem[{Erasmus {et~al.}(2017)Erasmus, Mommert, Trilling, Sickafoose, van
  Gend, \& Hora}]{Erasmus2017}
Erasmus, N., Mommert, M., Trilling, D.~E., {et~al.} 2017, The Astronomical
  Journal, 154, 162

\bibitem[{Florczak {et~al.}(1998)Florczak, Barucci, Doressoundiram, Lazzaro,
  Angeli, \& Dotto}]{Florczak1998}
Florczak, M., Barucci, M., Doressoundiram, A., {et~al.} 1998, Icarus, 133, 233

\bibitem[{Gaffey(1984)}]{Gaffey1984}
Gaffey, M.~J. 1984, Icarus, 60, 83

\bibitem[{Hsieh \& Jewitt(2006)}]{Hsieh2006}
Hsieh, H., \& Jewitt, D. 2006, Science, 312, 561

\bibitem[{Kim {et~al.}(2016)Kim, Lee, Park, Kim, Cha, Lee, Han, Chun, \&
  Yuk}]{Kim2016}
Kim, S.~L., Lee, C.~U., Park, B.~G., {et~al.} 2016, Journal of the Korean
  Astronomical Society, 49, 37

\bibitem[{Knezevic {et~al.}(2002)Knezevic, Lema{\^{i}}tre, \&
  Milani}]{Knezevic2002}
Knezevic, Z., Lema{\^{i}}tre, A., \& Milani, A. 2002, {The Determination of
  Asteroid Proper Elements}, ed. W.~{Bottke Jr.}, A.~Cellino, P.~Paolicchi, \&
  R.~Binzel No. 1979, 603--612

\bibitem[{Lazzaro {et~al.}(1999)Lazzaro, Moth{\'{e}}-Diniz, Carvano, Angeli,
  Betzler, Florczak, Cellino, {Di Martino}, Doressoundiram, Barucci, Dotto, \&
  Bendjoya}]{Lazzaro1999}
Lazzaro, D., Moth{\'{e}}-Diniz, T., Carvano, J., {et~al.} 1999, Icarus, 142,
  445

\bibitem[{Lomb(1976)}]{Lomb1976}
Lomb, N.~R. 1976, Astrophysics and Space Science, 39, 447

\bibitem[{Masiero {et~al.}(2015)Masiero, DeMeo, Kasuga, \&
  Parker}]{Masiero2015}
Masiero, J., DeMeo, F., Kasuga, T., \& Parker, A. 2015, {Asteroid Family
  Physical Properties}, ed. P.~Michel, F.~DeMeo, \& W.~Bottke, 323--340

\bibitem[{Milani {et~al.}(2014)Milani, Cellino, Kne{\v{z}}evi{\'{c}},
  Novakovi{\'{c}}, Spoto, \& Paolicchi}]{Milani2014}
Milani, A., Cellino, A., Kne{\v{z}}evi{\'{c}}, Z., {et~al.} 2014, Icarus, 239,
  46

\bibitem[{Mommert(2017)}]{Mommert2017}
Mommert, M. 2017, Astronomy and Computing, 18, 47

\bibitem[{Mommert {et~al.}(2016)Mommert, Trilling, Borth, Jedicke, Butler,
  Reyes-Ruiz, Pichardo, Petersen, Axelrod, \& Moskovitz}]{Mommert2016}
Mommert, M., Trilling, D.~E., Borth, D., {et~al.} 2016, The Astronomical
  Journal, 151, 98

\bibitem[{Nesvorny(2015)}]{Nesvorny2015b}
Nesvorny, D. 2015, NASA Planetary Data System, 234, EAR

\bibitem[{Nesvorny {et~al.}(2015)Nesvorny, Broz, \& Carruba}]{Nesvorny2015}
Nesvorny, D., Broz, M., \& Carruba, V. 2015, arXiv:1502.01628

\bibitem[{Parker {et~al.}(2008)Parker, Ivezi{\'{c}}, Juri{\'{c}}, Lupton,
  Sekora, \& Kowalski}]{Parker2008}
Parker, A., Ivezi{\'{c}}, {\v{Z}}., Juri{\'{c}}, M., {et~al.} 2008, Icarus,
  198, 138

\bibitem[{Rivkin \& Emery(2010)}]{Rivkin2010}
Rivkin, A., \& Emery, J. 2010, Nature, 464, 1322

\bibitem[{Russell {et~al.}(2012)Russell, Raymond, Coradini, Mcsween, Zuber,
  Nathues, Sanctis, Jaumann, Konopliv, Preusker, Asmar, Park, Gaskell, Keller,
  Mottola, Roatsch, Scully, Smith, Tricarico, Toplis, Christensen, Feldman,
  Lawrence, \& Mccoy}]{Russell2012}
Russell, C.~T., Raymond, C.~A., Coradini, A., {et~al.} 2012, Science, 336, 684

\bibitem[{Scargle(1982)}]{Scargle1982}
Scargle, J.~D. 1982, Astrophys. J., 263, 835

\bibitem[{Spoto {et~al.}(2015)Spoto, Milani, \&
  Kne{\v{z}}evi{\'{c}}}]{Spoto2015}
Spoto, F., Milani, A., \& Kne{\v{z}}evi{\'{c}}, Z. 2015, Icarus, 257, 275

\bibitem[{Vernazza {et~al.}(2008)Vernazza, Binzel, Thomas, DeMeo, Bus, Rivkin,
  \& Tokunaga}]{Vernazza2008}
Vernazza, P., Binzel, R., Thomas, C., {et~al.} 2008, Nature, 454, 858

\bibitem[{Zappala {et~al.}(1990)Zappala, Cellino, \& Farinella}]{Zappala1990}
Zappala, V., Cellino, A., \& Farinella, P. 1990, Astronomical Journal, 100,
  2030

\bibitem[{Ziffer {et~al.}(2011)Ziffer, Campins, Licandro, Walker, Fernandez,
  Clark, Mothe-Diniz, Howell, \& Deshpande}]{Ziffer2011}
Ziffer, J., Campins, H., Licandro, J., {et~al.} 2011, Icarus, 213, 538

\end{thebibliography}

\appendix 
\vspace*{\fill}
\section{All Data}
\vspace*{\fill}

\begin{longrotatetable}
\startlongtable 
\begin{deluxetable*}{|c|l|c|c|c|c|c|c|c|c|cccc|c|c|}
	\floattable

	\tabletypesize{\tiny}		
	\tablecaption{Observations and Results}
	\tablecolumns{16}
	\tablenum{1}
	\tablewidth{0pt}
	\tablehead{
		\colhead{No.} &\colhead{Object} &\colhead{Obs. Start}  &\colhead{Obs. Dur.} & \colhead{H\tablenotemark{\textit{a}}} & \colhead{V}& \colhead{V-R\tablenotemark{\textit{b}}}& \colhead{V-I\tablenotemark{\textit{b}}} & \colhead{Amplitude\tablenotemark{\textit{c}}} &\colhead{Rot. Period\tablenotemark{\textit{c}}}& \colhead{S-type}& \colhead{X-type}&\colhead{C-type}&\colhead{D-type}&\colhead{Tax.}&\colhead{Family\tablenotemark{\textit{d}}}\\
		\colhead{} &\colhead{} &\colhead{(JD)} & \colhead{(h:mm)} & \colhead{(mag)} &\colhead{(mag)} & \colhead{(mag)} & \colhead{(mag)}&  \colhead{(mag)} & \colhead{(min)}&  \multicolumn{4}{c}{(probability)}&\colhead{}& \colhead{}
	}
	\startdata
	\input{all_data_tex.txt}
	\enddata
	\footnote{}
	\tablenotetext{\textit{a}}{\textit{H} magnitude was obtained from \url{https://ssd.jpl.nasa.gov/horizons.cgi}}	
	\tablenotetext{\textit{b}}{Colors have been corrected for solar colors by subtracting the respective  \textit{V}$-$\textit{R} $=0.41$ and \textit{V}$-$\textit{I} $=0.75$ solar colors \citep{Binney1998}.}
	\tablenotetext{\textit{c}}{Lower limit shown in cases where observational duration was insufficient to observe entire light-curve period}
	\tablenotetext{\textit{d}}{We cross-correlate with NASA's Small Bodies Data Ferret database (SBDF) \citep{Nesvorny2015b} to associate objects from our survey data with known collisional families.}
	\label{table_all_data}
\end{deluxetable*}
\end{longrotatetable}
\end{document}